\documentclass[a4paper]{article}

\usepackage{icrc2013}
\usepackage[english]{babel}
\usepackage{amsmath,amssymb}
\title{Cosmic-ray spectral anomaly at GeV-TeV energies as due to re-acceleration by weak shocks in the Galaxy}

\shorttitle{Cosmic-ray spectral anomaly at GeV-TeV energies}

\authors{
Satyendra Thoudam$^{1,*}$ and
J\"org R. H\"orandel$^{1,2}$
}

\afiliations{
$^1$ Department of Astrophysics, IMAPP, Radboud University Nijmegen, 6500 GL Nijmegen, The Netherlands\\
$^2$ Nikhef, Science Park Amsterdam, 1098 XG Amsterdam, The Netherlands
}

\email{* s.thoudam@astro.ru.nl}

\abstract{Recent cosmic-ray measurements by the ATIC, CREAM and PAMELA experiments have found an apparent hardening of the energy spectrum at TeV energies. Although the origin of the hardening is not clearly understood, possible explanations include hardening in the cosmic-ray source spectrum, changes in the cosmic-ray propagation properties in the Galaxy and the effect of  nearby sources. In this contribution, we propose that the spectral anomaly might be an effect of re-acceleration of cosmic rays by weak shocks in the Galaxy. After acceleration by strong supernova remnant shock waves, cosmic rays undergo diffusive propagation through the Galaxy. During the propagation, cosmic rays may again encounter expanding supernova remnant shock waves, and get re-accelerated. As the probability of encountering old supernova remnants is expected to be larger than the young ones due to their bigger size, re-acceleration is expected to be produced mainly by weaker shocks. Since weaker shocks generate a softer particle spectrum, the resulting re-accelerated component will have a spectrum steeper than the initial cosmic-ray source spectrum produced by strong shocks. For a reasonable set of model parameters, it is shown that such re-accelerated component can dominate the GeV energy region while the non-reaccelerated component dominates at higher energies, explaining the observed GeV-TeV spectral anomaly.}

\keywords{cosmic rays, diffusion, supernova remnants.}

\begin{document}
\maketitle

\section{Introduction}
Recent measurements of cosmic rays by the ATIC \cite{bib:Panov2007}, CREAM \cite{bib:Yoon2011}, and PAMELA \cite{bib:Adriani2011} experiments have found a spectral anomaly at GeV-TeV energies. The spectrum in the TeV region is found to be harder than at  GeV energies. The spectral anomaly is difficult to explain under the standard models of cosmic-ray acceleration and their propagation in the Galaxy, which predict a single power-law spectrum over a wide range in  energy.

Various explanations for the spectral anomaly have been proposed. These include hardening in the cosmic-ray source spectrum at high energies \cite{bib:Biermann2010, bib:Ohira2011, bib:Yuan2011, bib:Ptuskin2013}, change in the cosmic-ray propagation properties in the Galaxy \cite{bib:Tomassetti2012, bib:Blasi2012}, and the effect of nearby sources \cite{bib:Thoudam2012, bib:Thoudam2013, bib:Erlykin2012, bib:Zatsepin2013}.

In this contribution, we discuss the possibility that the anomaly can be an effect of re-acceleration of cosmic rays by weak shocks in the Galaxy. This scenario was also shortly discussed recently by Ptuskin et al. 2011 \cite{bib:Ptuskin2011}. After acceleration by strong supernova remnant shock waves, cosmic rays escape from the remnants and undergo diffusive propagation in the Galaxy. The propagation can be accompanied by some level of re-acceleration due to repeated encounters with expanding supernova remnant shock waves \cite{bib:Wandel1988, bib:Berezhko2003}. As older remnants occupy a larger volume in the Galaxy, cosmic rays are expected to encounter older remnants more often than the younger ones. Thus, this process of  re-acceleration is expected to be produced mainly by weaker shocks. As weaker shocks generate a softer particle spectrum, the resulting re-accelerated component will have a spectrum steeper than the initial cosmic-ray source spectrum produced by strong shocks. As will be shown later, the re-accelerated component can dominate at GeV energies, and the non-reaccelerated component (hereafter referred to as the ``normal component") dominates at higher energies.   

Cosmic rays can also be re-accelerated by the same magnetic turbulence responsible for their scattering and spatial diffusion in the Galaxy. This process, which is commonly known as the distributed re-acceleration, has been studied quite extensively, and it is known that it can produce strong features on some of the observed properties of cosmic rays at low energies. For instance, the peak in the secondary-to-primary ratios at $\sim 1$ GeV/n can be attributed to this effect \cite{bib:Seo1994}. Earlier studies suggest that a strong amount of re-acceleration of this kind can produce unwanted bumps in the cosmic-ray proton and helium spectra at few GeV/n \cite{bib:Cesarsky1987, bib:Stephens1990}. However, it was later shown that for some mild re-acceleration which is sufficient to reproduce the observed boron-to-carbon ratio, the resulting proton spectrum does not show any noticeable bumpy structures \cite{bib:Seo1994}. In fact, the efficiency of distributed re-acceleration is expected to decrease with energy, and its effect becomes negligible at energies above $\sim 20$ GeV/n.

On the other hand, for the case of encounters with old supernova remnants mentioned earlier, the re-acceleration efficiency does not depend on energy. It depends only on the rate of supernova explosions and the fractional volume occupied by supernova remnants in the Galaxy. Hence, its effect can be extended to higher energies compared to that of the distributed re-acceleration, as also noted in Ref. \cite{bib:Ptuskin2011}. In the present study, we will first determine the maximum amount of re-acceleration permitted by the available measurements on the boron-to-carbon ratio. Then, we will apply the same re-acceleration strength to the proton and helium nuclei, and check if it can explain the observed spectral hardening for a reasonable set of model parameters.        

\section{Transport equation with re-acceleration}
Following Ref. \cite{bib:Wandel1987}, the re-acceleration of cosmic rays in the Galaxy is incorporated in the cosmic-ray transport equation as an additional source term with a power-law momentum spectrum. Then, the steady-state transport equation for cosmic-ray nuclei undergoing diffusion, re-acceleration and interaction losses can be written as,
\begin{align}
\nabla\cdot(D\nabla N)&-\left[\left(\bar{n} v\sigma+\xi\right)N+\xi sp^{-s}\int^p_{p_0}du\;N(u)u^{s-1}\right]\delta(z)\nonumber\\
&=-Q\delta(z)
\end{align}
where we use cylindrical spatial coordinates with the radial and vertical distance represented by $r$ and $z$ respectively, $p$ is the momentum/nucleon of the nuclei, $N(\textbf{r},p)$ represents the differential number density, $D(p)$ is the diffusion coefficient, and $Q(r,p)$ represents the source term. The first term on the left-hand side of Eq. (1) represents diffusion. The second and third terms represent losses due to inelastic interactions with the interstellar matter and due to re-acceleration to higher energies respectively, where $\bar{n}$ represents the averaged surface density of interstellar atoms, $v(p)$ the particle velocity, $\sigma(p)$ the inelastic collision cross-section, and $\xi$ corresponds to the rate of re-acceleration. The fourth term with the integral represents the generation of particles via re-acceleration of lower energy particles. It assumes that a given cosmic-ray population is instantaneously re-accelerated to form a power-law distribution with an index $s$. We neglect ionization losses and the effect of convection due to the Galactic wind as these processes are important mostly at energies below $\sim 1$ GeV/nucleon. The present study concentrates only at energies above $1$ GeV/nucleon.

The cosmic-ray propagation region is assumed as a cylindrical region bounded in the vertical direction at $z=\pm H$, and unbounded in the radial direction. Both the matter and the sources are assumed to be uniformly distributed in an infinitely thin disk of radius $R$ located at $z=0$. This assumption is based on the known high concentration of supernova remnants, and atomic and molecular hydrogens near the Galactic plane. For cosmic-ray primaries, the source term is taken as $Q(r,p)=\bar{\nu} Q(p)$, where $\bar{\nu}$ denotes the rate of supernova explosions (SNe) per unit surface area on the disk. The source spectrum is assumed to follow a power-law in total momentum with a high-momentum exponential cut-off. In terms of momentum/nucleon, it can be expressed as 
\begin{equation}
Q(p)=AQ_0 (Ap)^{-q}\exp\left(-\frac{Ap}{Zp_c}\right)
\end{equation}
where $A$ and $Z$ represents the mass number and charge of the nuclei respectively, $Q_0$ is a constant related to the amount of energy $f$ injected into a cosmic ray species by a single supernova event, $q$ is the source spectral index, and $p_c$ is the high-momentum cut-off for protons. In writing Eq. (2), we assume that the maximum total momentum (or energy) for a cosmic-ray nuclei produced by a supernova remnant is $Z$ times that of the protons. We further assume that the source spectrum has a low-momentum/nucleon cut-off at $p_0$ which also serves as the lower limit in the integral in Eq. (1). Moreover, the diffusion coefficient as a function of particle rigidity is assumed to follow  $D(\rho)=D_0\beta(\rho/\rho_0)^a$, where $\rho=Apc/Ze$ is the particle rigidity, $\beta=v/c$, and $c$ is the velocity of light.

In the present model, since the re-acceleration of cosmic rays is considered to be produced by their encounters with  supernova remnants, it follows that re-acceleration occurs only in the Galactic disk. If $V=4\pi \Re^3/3$ is the volume occupied by a supernova remnant of radius $\Re$, then in Eq. (1), $\xi=\eta V\bar{\nu}$, where $\eta$ is a correction factor for $V$ we have introduced to take care of the unknown actual volume of the supernova remnants that re-accelerate cosmic rays. We keep $\eta$ as a parameter, and we take $\Re=100$ pc which is roughly the typical radius of a supernova remnant of age $10^5$ yr expanding in the interstellar medium with an initial shock velocity of $10^9$ cm s$^{-1}$. 

The solution of Eq. (1) is obtained using the standard Green's function technique. For sources uniformly distributed in the Galactic disk, the solution at $r=0$ is obtained as,
\begin{align}
N(z,p)&=\bar{\nu} R\int^{\infty}_0 dk\; \frac{\sinh\left[k(H-z)\right]}{\sinh(kH)}\times \frac{\mathrm{J_1}(kR)}{L(p)}
\times F(p) 
\end{align}
where $\mathrm{J_1}$ is a Bessel function of order 1, 
\begin{equation}
L(p)=2D(p)k\coth(kH)+\bar{n}v(p)\sigma(p)+\xi,
\end{equation}
\begin{eqnarray}
F(p)=Q(p)+\xi sp^{-s}\int^p_{p_0}u^s du\;Q(u)A(u)\nonumber\\
\times\exp\left(\xi s\int^p_u A(w)dw\right) 
\end{eqnarray}
and, the function $A$ is given by
\begin{equation}
A(x)=\frac{1}{xL(x)}
\end{equation}

Considering that the position of our Sun is very close to the Galactic plane, the cosmic-ray density at the Earth can be calculated from Eq. (3) taking $z=0$.

The first term on the right-hand side of Eq. (5) is the normal cosmic-ray component which have not suffered re-acceleration, and the second term is purely the re-accelerated component. For a given diffusion index, the high-energy spectra of the two components are shaped by their respective injection indices $q$ and $s$. As re-acceleration takes out particles from the low-energy region and puts them into the higher energy part of the spectrum, for re-acceleration by weak shocks for which $s>q$, the re-accelerated component might become visible as a bump or enhancement in the energy spectrum at a certain energy range. In the case of re-acceleration by strong shocks which produces a harder particle spectrum, say $s=q$, the effect of re-acceleration will be hard to notice as both the components will have the same spectra in the Galaxy. These have been extensively discussed in Ref. \cite{bib:Wandel1987}.  

For cosmic-ray secondaries, their equilibrium density $N_2(\textbf{r},p)$ in the Galaxy is obtained following the same procedure as for their primaries described above, but with the source term replaced by
\begin{equation}
Q_2(\textbf{r},p)=\bar{n} v_1(p)\sigma_{12}(p)N_1(\textbf{r},p) \delta(z)
\end{equation}
where $v_1$ represents the velocity of the secondary nuclei, $\sigma_{12}$ represents the total fragmentation cross-section of the primary to the secondary, and $N_1$ is the primary nuclei density. The subscripts $1$ and $2$ have been introduced to denote primary and secondary nuclei respectively.

The secondary-to-primary ratio can be calculated by taking the ratio $N_2/N_1$. For the case of no re-acceleration $\xi=0$, it can be checked that Eq. (3) reduces to the standard solution of pure-diffusion equation (see e.g., \cite{bib:Thoudam2008}), and also that the secondary-to-primary ratio becomes inversely proportional to the diffusion coefficient at high energies. It can be mentioned that a steeper re-acceleration index $s>q$ will produce an enhancement in the ratio at lower energies, and a harder index $s=q$ will result into significant flattening of the ratio at high energies \cite{bib:Berezhko2003, bib:Wandel1987}. Thus, the effect of re-acceleration on cosmic-ray properties in the Galaxy depends strongly on the index of re-acceleration. In the present study, since we assume that re-acceleration is produced mainly by the interactions with old supernova remnants, we will only consider the case of $s>q$ with $s\gtrsim 4$. This value of $s$ corresponds to a Mach number of $\sim 1.7$ of the shocks that re-accelerate the cosmic rays.

\section{Results and discussions}
For the interstellar matter density, we consider the averaged surface density on the Galactic disk within a radius equivalent to the halo height $H$. We take $H=5$ kpc for our study, and the averaged surface density of atomic hydrogen as $\bar{n}=7.24\times 10^{20}$ atoms cm$^{-2}$ \cite{bib:Thoudam2013}. We assume that the interstellar medium consists of $10\%$ helium. The inelastic interaction cross-sections are taken as the same used in the calculation in Ref. \cite{bib:Thoudam2013}. 

We take the size of the source distribution $R=20$ kpc, the proton low and high-momentum cut-offs as $p_0=100$ MeV/c and $p_c=1$ PeV/c respectively, and the supernova explosion rate as $\bar{\nu}=25$ SNe Myr$^{-1}$ kpc$^{-1}$. The latter corresponds to a rate of $\sim 3$ SNe per century in the Galaxy. The cosmic-ray propagation parameters $(D_0, \rho_0, a)$, the re-acceleration parameters $(\eta, s)$ and the source parameters $(q, f)$ are taken as model parameters. They are determined based on the measured data.

\begin{figure}
\includegraphics*[width=0.47\textwidth]{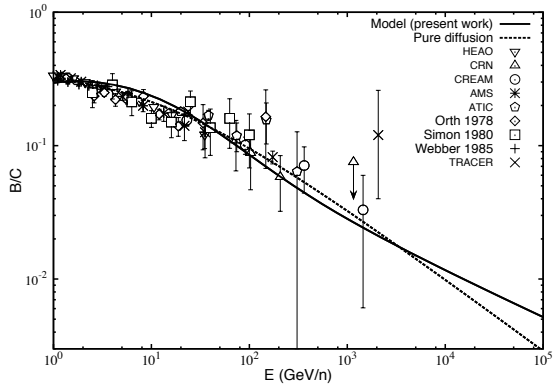}
\vspace{-5 mm}
\caption{Boron-to-Carbon (B/C) ratio. \textit{Solid line}: Present work including re-acceleration. \textit{Dashed line}: Pure diffusion model without re-acceleration \cite{bib:Thoudam2013}.}
\end{figure}

We first determine $(D_0, \rho_0, a, \eta, s)$ based on the measurement data for the boron-to-carbon ratio, and the spectra for the  carbon, oxygen, and boron nuclei. Their values are found to be $D_0=9\times 10^{28}$ cm$^2$ s$^{-1}$, $\rho=3$ GV, $a=0.33$, $\eta=1.02$, $s=4.5$. These values correspond to the maximum amount of re-acceleration permitted by the available boron-to-carbon data, while at the same time produces a reasonable good fit to the measured primary and secondary spectra. Figure 1 shows the result on the boron-to-carbon ratio (solid line) along with the measurement data. For comparison, we have also shown the result for the case of pure diffusion (dashed line) with no re-acceleration $(\eta=0)$ taken from Ref. \cite{bib:Thoudam2013}. The good fit carbon and oxygen source parameters are found to be $q_C=2.24, f_C=0.024\%$, and $q_O=2.26$, $f_O=0.025\%$ respectively, where the $f$'s are given in units of $10^{51}$ ergs. The present calculation assumes a force-field solar modulation parameter of $\phi=450$ MV.

\begin{figure}
\includegraphics[width=0.47\textwidth]{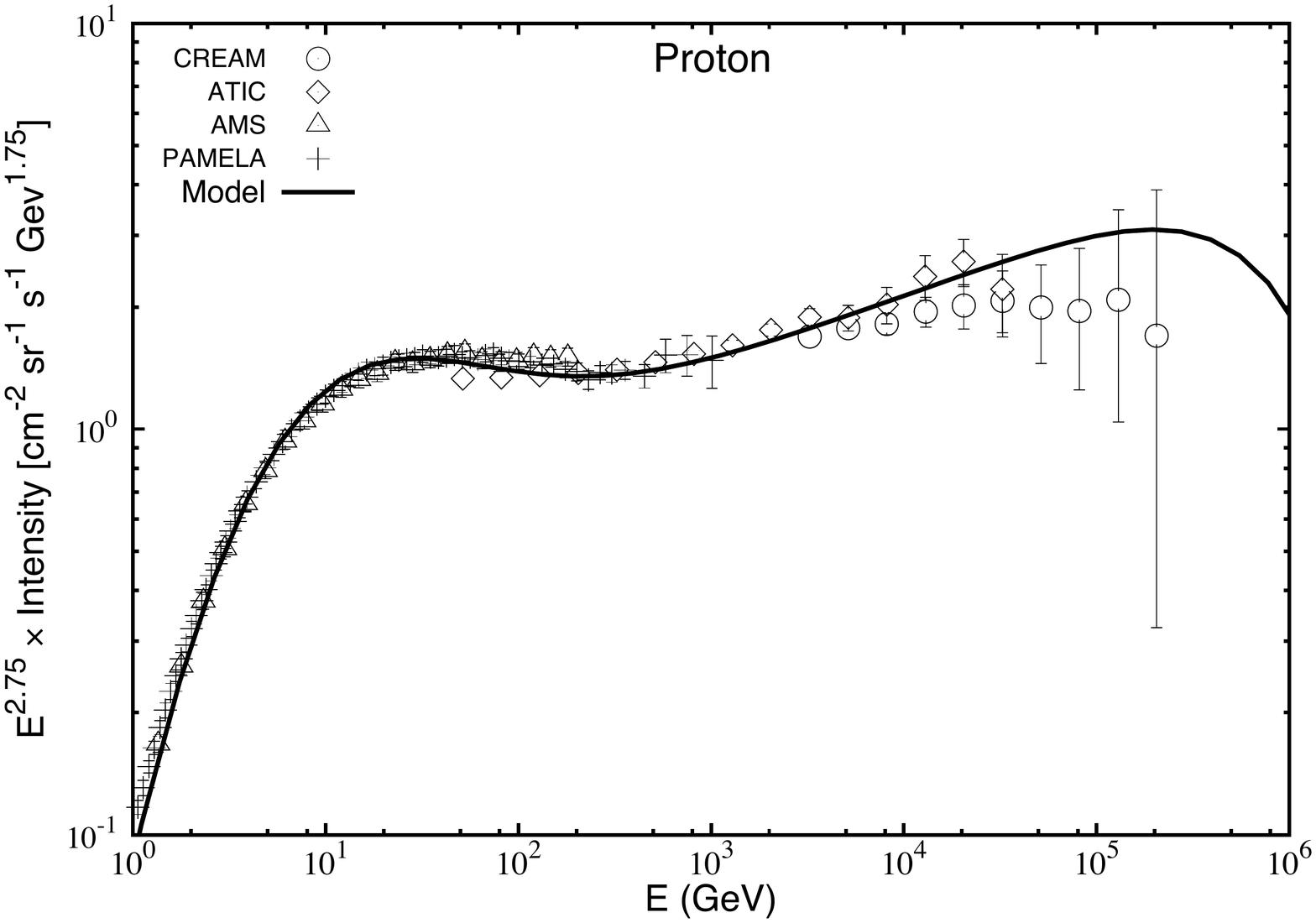}
\vspace{-7 mm}\\
\includegraphics[width=0.47\textwidth]{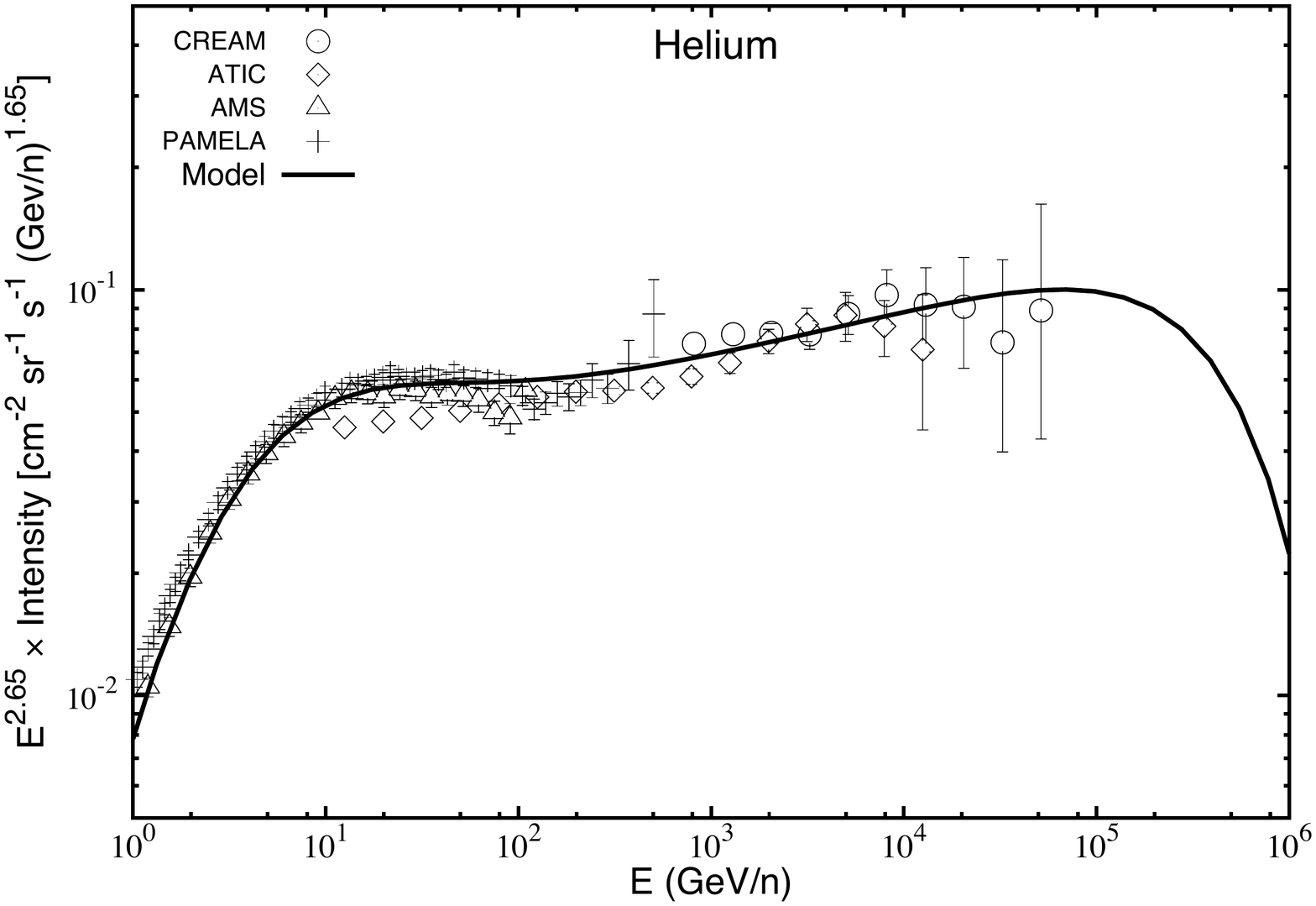}
\vspace{-5 mm}
\caption{\textit{Top}: Proton spectrum. \textit{Bottom}: Helium spectrum. The lines represent our results. For the data, see the experiments listed in Ref. \cite{bib:Thoudam2013}.}
\end{figure}

Using the same values of $(D_0, \rho_0, a, \eta, s)$ obtained above, we calculate the spectra for the proton and helium nuclei. The results are shown in Figure 2, where the top panel represents proton and the bottom panel represents helium. The lines represent our results, and the data are the same as used in Ref. \cite{bib:Thoudam2013}. The source parameters used are $q_p=2.21, f_p=6.95\%$ for protons, and $q_{He}=2.18, f_{He}=0.79\%$ for helium, and we use the same solar modulation parameter as given above. It can be seen that our results are in good agreement with the measured data, and explain the observed spectral anomaly between the GeV and TeV energy regions. Below $\sim 200$ GeV/n, our model spectrum is dominated by the re-accelerated component while above, it is dominated by the normal component.

\begin{figure}
\includegraphics*[width=0.47\textwidth]{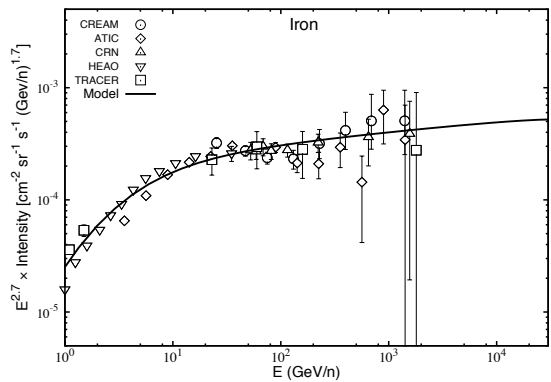}
\vspace{-5 mm}
\caption{Iron spectrum. The line represents our result. For the data, see the experiments listed in Ref. \cite{bib:Thoudam2013}.}
\end{figure}

The effect of re-acceleration is stronger in the case of protons than helium which is due to the larger inelastic collision losses for helium. This result into more prominent spectral differences in the GeV-TeV region for protons than for helium. For heavier nuclei for which the inelastic cross-sections are much larger, the re-acceleration effect is expected to be negligible. Figure 3 shows our result for iron nuclei. The calculation assumes $q_{Fe}=2.28$, and $f_{Fe}=4.9\times 10^{-3}\%$. As expected, the re-acceleration effect is hard to notice in Figure 3, and the model spectrum above $\sim 20$ GeV/n follow approximately a single power-law unlike the proton and helium spectra. 

\section{Conclusions}
We have shown that the spectral anomaly at GeV-TeV energies, observed for the proton and helium nuclei, can be an effect of re-acceleration by weak shocks associated with old supernova remnants in the Galaxy. The re-acceleration effect is shown to be important  for light nuclei, and negligible for heavier nuclei such as iron. Our prediction of decreasing effect of re-acceleration with the elemental mass can be checked by future sensitive measurements of heavier nuclei at TeV/n energies. 


\end{document}